\documentstyle[11pt,epsfig,epsf]{article} 
\def\baselinestretch{1.3}
\newcommand{\ba}{\begin{array}}
\newcommand{\ea}{\end{array}}
\newcommand{\bd}{\begin{displaymath}}
\newcommand{\ed}{\end{displaymath}}
\newcommand{\be}{\begin{equation}}
\newcommand{\ee}{\end{equation}}
\newcommand{\bea}{\begin{eqnarray}}
\newcommand{\eea}{\end{eqnarray}}

\def\q2 {q^2}

\def\r {\rightarrow}

\def\slash {\!\!\!\!\!\!/}

\begin{document}
\begin{flushright}
{\large MRI-P-011005}\\ 
{hep-ph/0111012}\\
\end{flushright}

\begin{center}
{\Large\bf Invisible  charginos and neutralinos from gauge boson fusion : \\ a way to explore
 anomaly mediation ?}\\[20mm]
Anindya Datta \footnote{E-mail: anindya@mri.ernet.in},
Partha Konar \footnote{E-mail: konar@mri.ernet.in}   and 
Biswarup Mukhopadhyaya \footnote{E-mail: biswarup@mri.ernet.in}\\
{\em Harish-Chandra Research Institute,\\
Chhatnag Road, Jhusi, Allahabad - 211 019, India}
\end{center}

\vskip 20pt
\begin{abstract}
  We point out that vector boson fusion (VBF) at the Large Hadron
  Collider (LHC) can lead to useful signals for charginos and
  neutralinos in supersymmetric scenarios where these particles are
  almost invisible. The proposed signals are just two forward jets
  with missing transverse energy. It is shown that in this way one can
  put by far the strongest constraint on the parameter space of a 
  theory with anomaly
  mediated supersymmetry breaking (AMSB) at the LHC. In addition,
  scenarios where the lightest neutralinos and charginos are
  Higgsino-like can give signals of the above type.
\end{abstract}

\vskip 1 true cm

\newpage
\setcounter{footnote}{0}

\def\baselinestretch{1.8}

Vector boson fusion (VBF) at hadronic machines such as the Large
Hadron Collider (LHC) at CERN has been suggested as a useful channel
for studying the signal of the Higgs boson. Characteristic features of
this mechanism are two highly energetic quark-jets, produced in the
forward direction in opposite hemispheres and carrying a large
invariant mass.  The absence of colour exchange between the forward
jets ensures a suppression of hadronic activities in the central
region \cite{bjorken}. Though it was originally proposed as a
background-free signal of a heavy Higgs \cite{dawson}, the usefulness
of the VBF channel in uncovering an intermediate mass Higgs has also
been subsequently demonstrated \cite{zeppenfeld}.

Encouraged by all this, one naturally wants to know whether the VBF
channel can be used to unravel other aspects of the basic constituents
of nature, especially those bearing the stamp of physics beyond the
standard model of elementary particles. It should be emphasized here
that the tagging of forward-jet events is part of the experimental
programme at the LHC, and therefore any new physics contributing to
such events is bound to get explored there.

One such candidate scenario, constantly knocking at our door, is
supersymmetry (SUSY). While a multitude of signals for SUSY at the
up-and-coming accelerators have been proposed \cite{susy_search}, here
we want to stress the utility of VBF processes to probe the
non-strongly interacting sector of the SUSY standard model. We point
out in particular that some of the SUSY theories of considerable
current interest can be tested in this way, via signals where nothing
excepting the forward-tagged jets are visible.

When R-parity (defined as $R =(-1)^{3B+L+2S}$) is conserved, a
conventional method of searching for charginos ($\chi^{\pm}$) and
neutralinos ($\chi^0$) at hadron colliders is their direct production.
The most convenient channel is $p \bar{p}~/~pp ~ \r ~ \chi_1^{\pm}
~\chi_2^{0}$ followed by the decays $\chi_1^{\pm} ~ \r ~ \chi_1^{0}
l^{\pm} {\nu_l} (\bar{\nu_l})$ and $\chi_2^{0} ~ \r ~ \chi_1^{0} l^{+}
l^{-}$, where $\chi^{0}_1$ is the lightest SUSY particle (LSP) and
hence is invisible.  This gives rise to `hadronically quiet' trilepton
signals \cite{hadq}.

It is in cases where the trilepton signal is not expected to be
visible that other channels such as VBF must be explored, if one wants
to study the non-strongly interacting sector in isolation. For
example, in some of the currently popular SUSY models, the lighter
chargino ($\chi^{\pm}_1$) and the lightest neutralino ($\chi^{0}_1$)
are closely degenerate in mass. Then the previously mentioned
trilepton signal is no more detectable, since the chargino decays into
either a soft pion ($\pi$) or very soft leptons/quarks together with
the $\chi^{0}_1$.  This makes the chargino-neutralino pair essentially
invisible. Final states comprising them need to be identified with
some visible tags. In electron-positron colliders, the use of a photon
as such a tag is advocated \cite{photon_tag}. However, tagging either
photons or gluons at hadronic machines is unlikely to be efficient due
to extremely large backgrounds. Under the circumstances, we find it
useful to study the production of chargino-neutralino pairs in VBF,
since the forward jets themselves act as the necessary tags.  In such
cases, two $ forward~jets + E_T \slash ~$ can be treated as the generic
signal of invisibly decaying charginos and neutralinos.

It should be noted that such signals have already been suggested
\cite{eboli} for Higgs bosons decaying invisibly into stable, neutral
weakly interacting particles like a pair of LSP's in the minimal SUSY
standard model (MSSM), or pairs of gravitinos or Majorons in some other
extended theories.  It has also been shown in the above reference that
the backgrounds to such a signal can be effectively handled on using 
suitable event selection criteria.  We demonstrate that the partial 
invisibility of the chargino-neutralino sector of a SUSY model can be 
used to our advantage using very similar event selection criteria.

The effectiveness of the VBF technique has also been demonstrated by
us in an R-parity violating scenario where it is rather difficult to
distinguish the final states obtained via $p \bar{p}~/~pp ~ \r ~
\chi_1^{\pm} ~\chi_2^{0}$ against cascades coming from the strongly
interacting sector \cite{our_pap}. In VBF channel, charginos and
neutralinos produced with the help of the $W$-boson, the $Z$-boson and
the photon lead to signatures of such models in the form of like-and
unlike-sign dileptons in background-free environments.

We will consider two specific examples which are of interest for the
present purpose.  The first one of these is a theory with Anomaly
Mediated Supersymmetry Breaking (AMSB) \cite{amsb} where
$\chi^{\pm}_1$ $\chi^0_1$ are both wino-like and therefore very
closely degenerate.  The second instance is that of a SUSY Grand
Unified Theory (GUT) with $M_2 >> \mu$, where $M_2$ is the $SU(2)$ 
gaugino mass and $\mu$, the Higgsino mass parameter occurring
in the superpotential. In that
case, $\chi^0_1$, $\chi^{\pm}_1$ and $\chi^0_2$ are all Higgsino-like
and have small mass separations.  We show below how it is possible to
constrain both of these scenarios at the LHC using the VBF mechanism.

AMSB models attempt to link the SUSY breaking mechanism to scenarios
with extra compactified dimensions. The SUSY breaking sector
is confined to a  3-brane separated from the one on which the standard
model fields reside. SUSY breaking is conveyed to the observable
sector by a super-Weyl anomaly terms, and the gaugino and scalar 
masses are given by

\bea 
M_i &=& b_i\;\frac{g_i^2}{16 \pi ^2}\;m_{3/2} \nonumber \\
M^2_{scalar} &=& c^2\;\frac{m_{3/2}^2}{(16 \pi^2) ^2}\;+\;m_0^2 
\eea

Here $b_i$s' are coefficients occurring in the $\beta$-functions of the
appropriate gauge couplings and $c$'s are combinations of
$\beta$-functions and anomalous dimensions (of gauge and Yukawa
couplings).  Explicit expressions for these can be obtained, for example, 
from \cite{sourov}. $m_0$ is a scalar mass parameter introduced to prevent
sleptons from becoming tachyonic.

\begin{figure}
\centerline{
\epsfxsize= 7 cm\epsfysize=7.0cm
                     \epsfbox{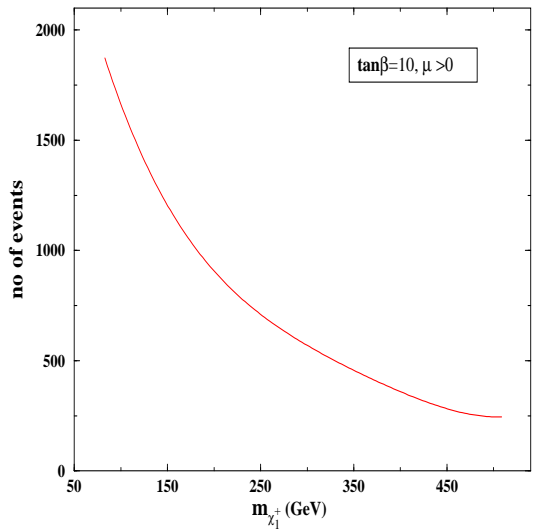}
}
 
\caption{{\it Variation of number of (2 forward jets + $ E_T\slash~$) events (after applying
   cuts as specified in the text) with 
    the lighter chargino mass ($m_{\chi^+_1}$) in AMSB with $\tan\beta
    = 10$ and $\mu > 0$. An integrated luminosity of $100~fb^{-1}$
    has been assumed. }}
\label{fig1}
\end{figure}

Since the gaugino masses are proportional to the beta-functions of the
corresponding gauge couplings, both the lightest neutralino (which is
the LSP) and the lighter chargino turn out to be dominated by the
wino, with their masses separated by a few hundreds of $MeV$.  The
second lightest neutralino, on the hand hand, is about three times
larger in mass and is Bino-dominated.

This kind of a spectrum implies that the dominant decay mode for
the lighter chargino is $\chi^{\pm}_1 \longrightarrow \pi^{\pm}
\chi^{0}_1$. The pion in such cases is too soft to be detected, making
the chargino essentially invisible. There are suggested ways of
looking for this kind of a spectrum in high-energy $e^{+}e^{-}$
colliders \cite{photon_tag, sourov}. Studies to probe AMSB at hadron
colliders using different superparticle decay cascades are also found
in the literature \cite{amsb_tata}. We, on the other hand, exploit the
invisibility of the lighter chargino and tagging of the forward jets
in the VBF channel to explore or exclude this kind of a theory at the
LHC.  Our analysis is based on the processes $pp \longrightarrow
\chi^{\pm}_1 \chi^{0}_1~jj$, $pp \longrightarrow \chi^{+}_1
\chi^{-}_1~jj$ and $pp \longrightarrow \chi^{0}_1 \chi^{0}_1~jj$,
driven by the fusion of gauge bosons, which give rise
to just two visible forward jets with missing transverse energy.
Similar final states may also arise from $\chi^{\pm}_1 \chi^{0}_2$
production, but the contribution to the events of our interest is
small, since $(a)$ the Bino-dominated character of $\chi^{0}_2$ makes
the production rate low, and $(b)$ the invisible final states can only
arise from $\chi^{0}_2 \longrightarrow \nu {\bar{\nu}} \chi^{0}_1$
where a further suppression by the branching fraction takes place.

The signals, however, are not background-free. As has been
already discussed in reference \cite{eboli}, such events can be faked
by the pair-production of neutrinos along with two forward jets. In
addition, two forward jets together with a soft lepton and missing
$E_T$ (due to a neutrino) can also fake the signal. Such final states 
can arise 
in the standard model from real emission corrections to the Drell-Yan 
process as well as from electroweak $W$ and $Z$ production along 
with two jets.

Keeping all these in mind, we have applied the following event selection
criteria:

\begin{itemize}
\item Two forward jets in opposite hemispheres, 
      with $E_T > 40~GeV$ and 
      $2.0 \le |\eta_j| \le 5.0$. 

\item $\Delta \eta_{jj} > 4$.

\item $M_{inv} (jj) > 1200~GeV$.

\item  ${E_T}\slash > 100~GeV$.

\item $\Delta \phi_{jj} < 57^o$.
\end{itemize}

\begin{figure}
\centerline{
\epsfxsize= 7 cm\epsfysize=6.5cm
                     \epsfbox{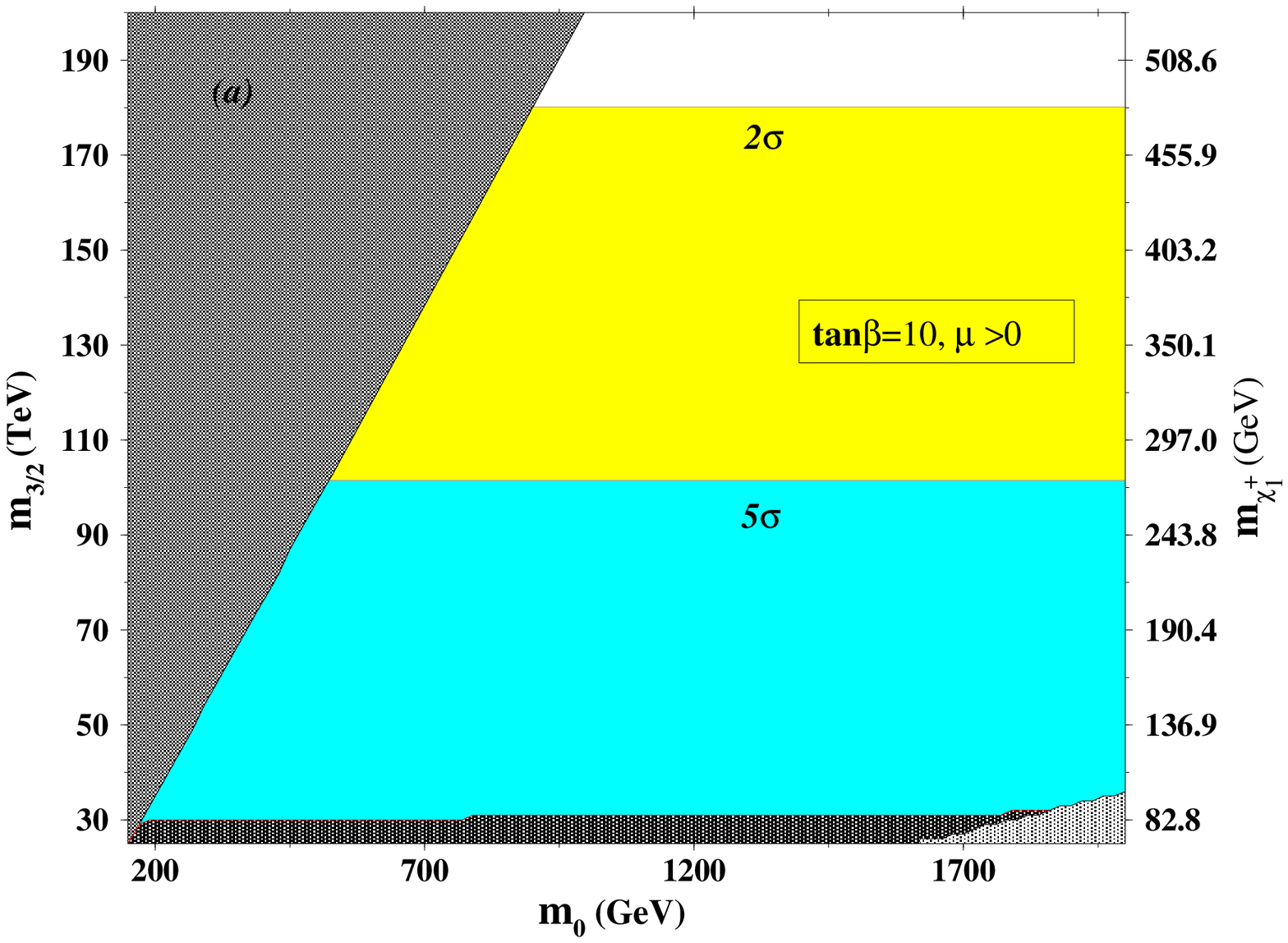}
        \hspace*{.7cm}
\epsfxsize=7.0 cm\epsfysize=6.5cm
                     \epsfbox{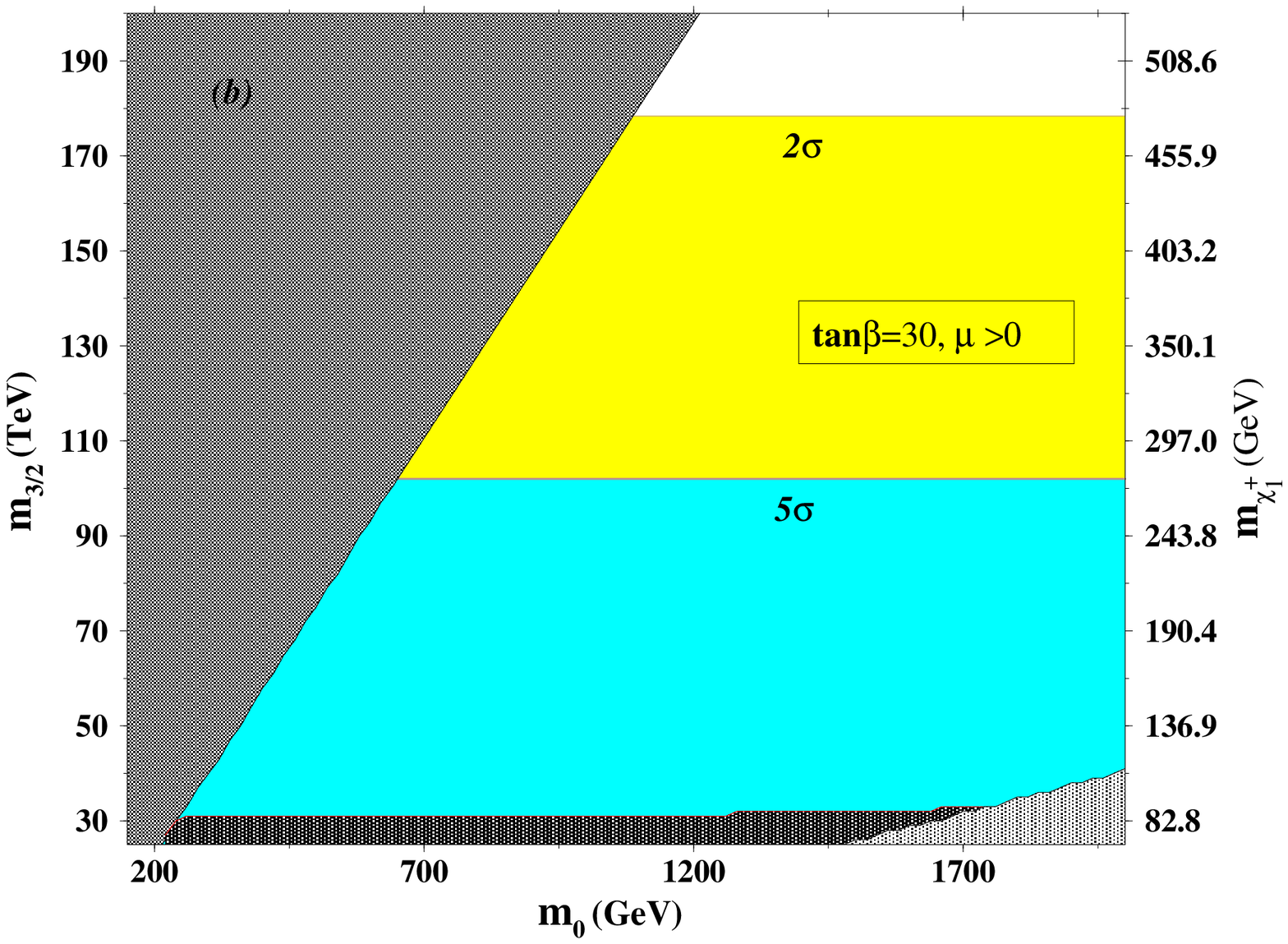}
}
 
\caption{{\it 5$\sigma$ and 2$\sigma$ discovery regions in the $m_0 - m_{3/2}$ plane for
    $\mu > 0$ and (a) $tan\beta = 10$ and (b) $tan\beta = 30$ in an
    AMSB scenario. The upper-left shaded region is excluded to prevent
    the lighter $\tilde \tau$ becoming first the LSP and then tachyonic. 
    The dark shaded region parallel to the $m_0$ axis in low $m_{3/2}$ 
    is disallowed from LEP data. The light shaded portion in the
    lower-right corner is excluded to ensure electroweak symmetry
    breaking.  }}
\label{fig2}
\end{figure}

The first three cuts establish the {\em bona fide} of the VBF events.
All the criteria enhance the signal-to-background ratio, since the
background events are in general found to have smaller missing $E_T$
and tend to have back-to-back orientations in the transverse
plane. Also, we have required the pions to have $E_T$ less than
$20~GeV$ to be really undetected. The leading order estimate of
backgrounds depends on the choice of the renormalization
scale. Estimates with different choices are found in reference
\cite{eboli}.  The range over which such estimates vary in our case,
with and without the rapidity and azimuthal angle cuts, are shown in
table 1.  The purpose of this table is to show the dependence on
renormalization scale choice and the effects of kinematic cuts
(especially the lower cut on rapidity and the upper cut on the
azimuthal angle). Both the azimuthal angle cut and the lower cut on
jet rapidity cause a substantial reduction on the background rate. As
has been already mentioned, the backgrounds can have their origin in
both QCD and electroweak interactions. For the choice of higher
$\alpha_s$ in table 1, for example, one has 168 $fb$ of QCD
contribution, and 21 $fb$ from electroweak processes. 
The numbers in the first column of the table are obtained when 
all the cuts specified earlier in the text are applied except the 
one on $\Delta \phi _{jj}$. Similarly, for the second column,
we only remove the lower cut on the rapidity of the forward jets
($| \eta _j | > 2$). Therefore, a comparison of the numbers 
in the first two columns with the corresponding entries in the third
column also give us an estimate of the efficiencies of these cuts
individually. Furthermore, one multiplies the cross-sections with 
the survival rates on the application of the central jet veto. the 
survival probabilities have been taken as 0.9, 0.28 and 0.82 respectively 
for the signal, QCD backgrounds and electroweak backgrounds \cite{rain}.

The amplitudes for all the processes of our interest have been
calculated using the HELAS helicity amplitude subroutines
\cite{helas}. We have used CTEQ4L parton distribution functions 
\cite{cteq}.

In figure 1 we plot the number of signal events, calculated for an
integrated luminosity of $100~fb^{-1}$ at the LHC, against the
chargino mass for $\mu~>~0$ and $\tan \beta~=~10$. The rates,
essentially dependent on the chargino mass, are by and large
insensitive to $\tan \beta$ and $m_0$.

In estimating the detectability of the signal against the backgrounds, 
we have chosen a renormalization
scale that keeps $\alpha_s$ on the high side. Thus we have taken
$\alpha_s=\alpha_s(\,min \{{p_T}_{j_1},{p_T}_{j_2}\}\,)$. 
Performing an analysis similar to that in 
\cite{eboli}, we have utilized the fact that the backgrounds can be 
estimated to a fairly high precision ($\sim 1.2\%$) at the high luminosity 
option of the LHC, from the visible decay products of the W and the Z. 
The above uncertainty has been added in quadrature
with the Gaussian fluctuation in the background itself. 
It is then possible to identify those regions of the parameter space 
where the signal survives at difference confidence levels.

\begin{table}
\begin{center}
\begin{tabular}{||l|c|c|c||}    \hline\hline
\emph{choice of the }&
\multicolumn{3}{c||}{\emph{Background (in fb)}}\\ \cline{2-4}
\emph{renormalization scale} & \emph{without $\Delta \phi_{jj} < 57^o$ cut }& 
\emph{without $|\eta_j|\ge 2 $ cut } &  \emph{with all cuts}\\

\hline

$\alpha_s=\alpha_s(\,min \{{p_T}_{j_1},{p_T}_{j_2}\}\,)$& 984 & 452 & 189 
\\ \hline

$\alpha_s=\alpha_s(\sqrt{\hat{s}/4}\,)$ & 334 & 193  & 72 \\ \hline\hline
\end{tabular}
\end{center}
\caption{{Backgrounds for different choices of the renormalization scales
with and without the rapidity and azimuthal angle cuts (the remaining cuts
as specified in the text are retained in each case).}}
\label{table1}
\end{table}

It may be mentioned that another way of handling the backgrounds in
such a case has been suggested in the literature \cite{feng}. This consists
in identifying the track left by a possibly long-lived chargino. 
However, further studies are required to ascertain whether this is a 
viable option at the high-luminosity version of the LHC.

Figures 2(a) and 2(b) show the regions in the $m_0 - m_{3/2}$ plane
corresponding to $2\sigma$  and $5\sigma$
detectability of the signal against the backgrounds estimated above.
Again, an integrated luminosity of  $100~fb^{-1}$ has been assumed.  
We also show the regions already excluded by LEP data
\cite{lep_data} as well as those forbidden by the possibility of
tachyonic sleptons (or $\tilde{\tau}$-LSP) and the
impossibility of electroweak symmetry breaking. The signal event rates
are independent of $m_0$ and highly insensitive to $\tan \beta$, as
is evident on a comparison of the two figures.

It is clear from both the figures that 
the entire AMSB parameter space for $m_{3/2} \le 190~TeV$ 
can be probed at
95\%  confidence level, corresponding to a lighter chargino of mass upto
about $500~GeV$. Moreover, chargino masses upto $300~GeV$ can be explored 
at the $5\sigma$ level.       
The results are very similar for negative values of $\mu$.
The above predictions, it should be emphasized, correspond to
a case where backgrounds are assumed to be of the largest possible
magnitude. 
Therefore, {\em although our predictions are related to processes involving
charginos and neutralinos only, the signals to be looked for can have
wider implications, since the fundamental parameters
of AMSB can be constrained through them}.

\begin{figure}
\centerline{
\epsfxsize= 7 cm\epsfysize=7.0cm
                     \epsfbox{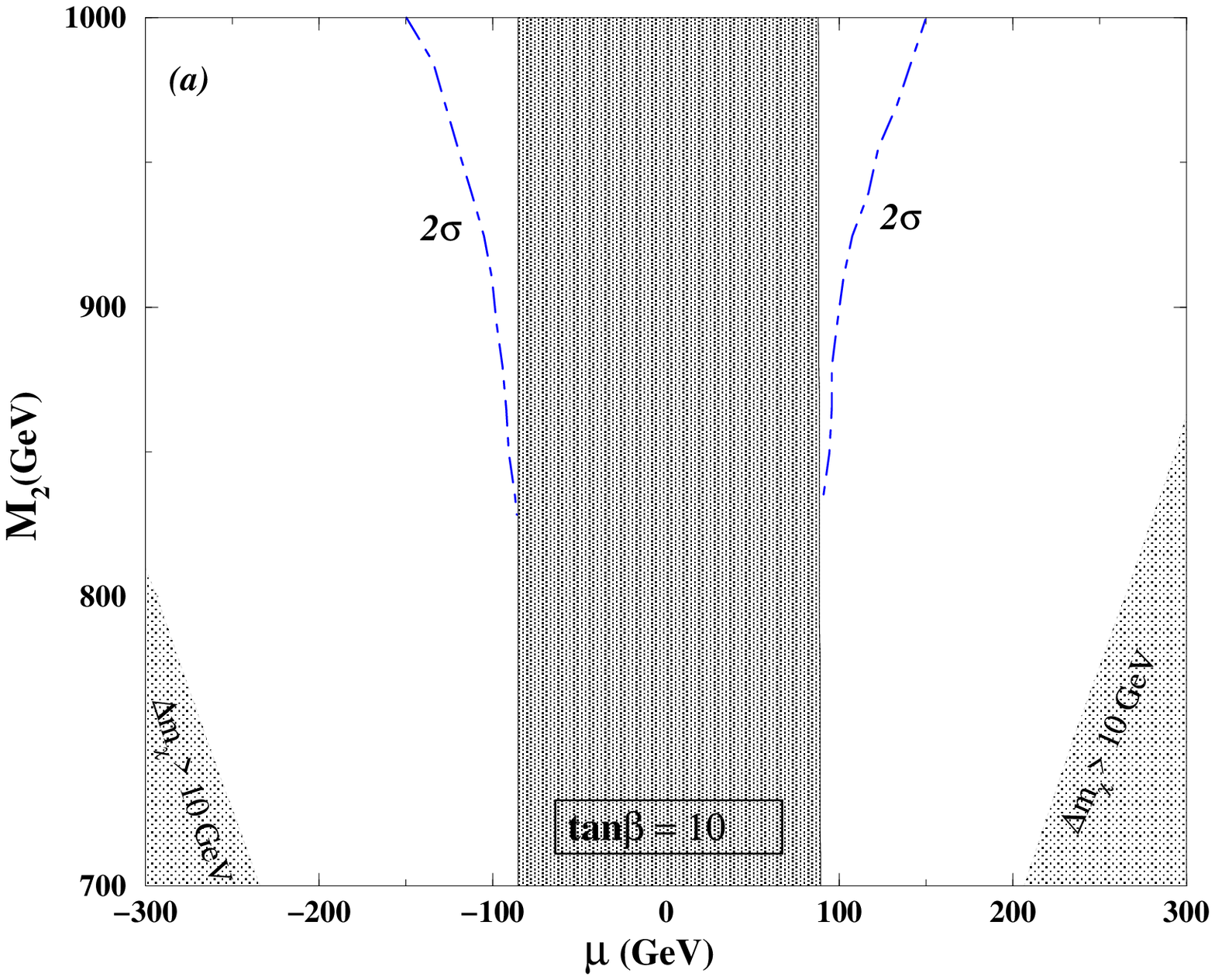}
        \hspace*{.7cm}
\epsfxsize=7.0 cm\epsfysize=7.0cm
                     \epsfbox{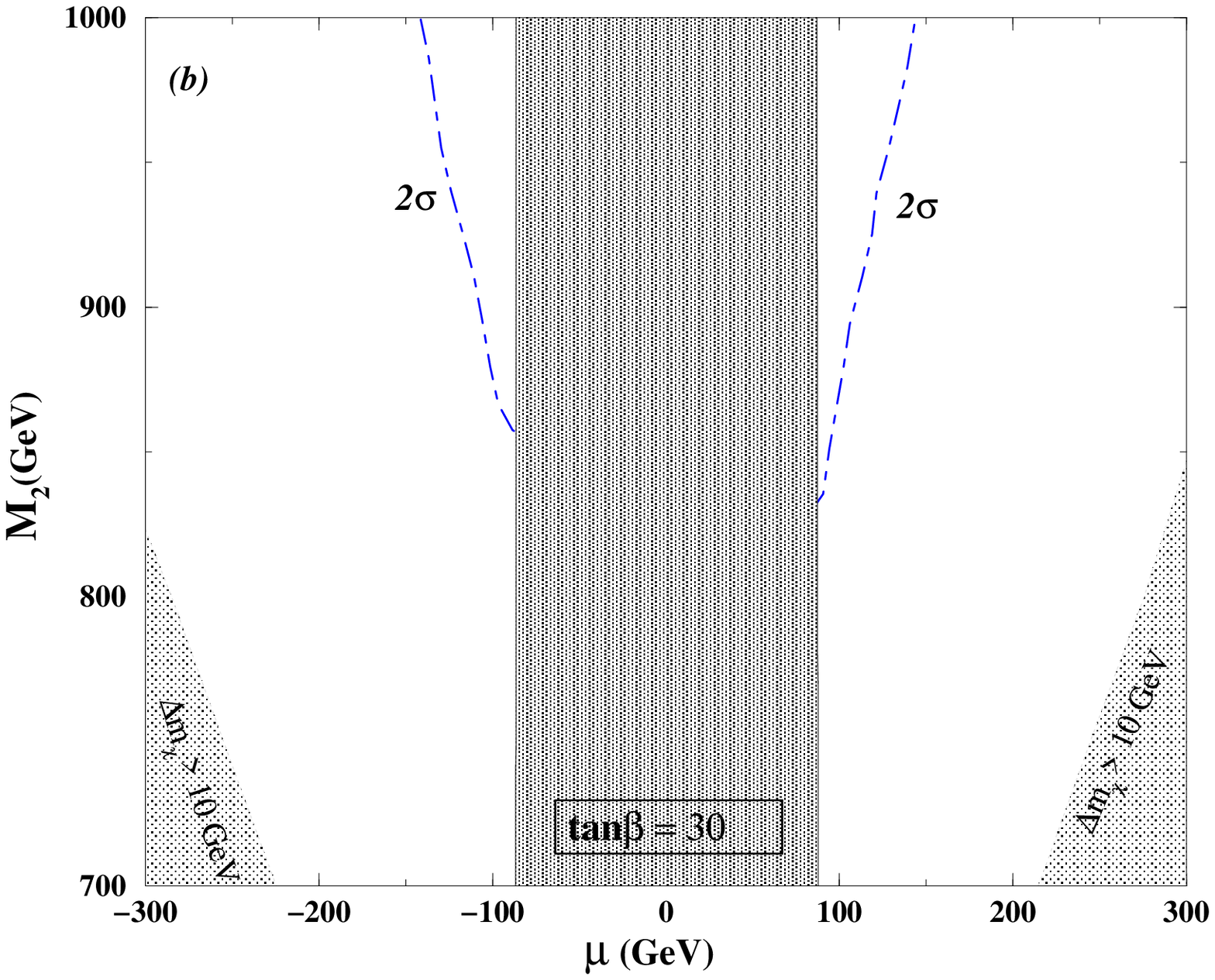}
}
 
\caption{{\it 2$\sigma$ significance contours in a SUSY GUT scenario
    for (a) $tan\beta = 10$ (b) $tan\beta = 30$ in the $\mu- M_2$
    plane. The dark shaded region is disallowed from LEP data. The
    shaded portions in the two corners indicate the regions where
    $(m_{\chi^{\pm}_1} - m_{\chi^{0}_1}) > 10~GeV$ }}
\label{fig3}
\end{figure}

It may be noted that general strategies for AMSB search at hadronic machines 
have been reported in earlier works (for example, in the first
reference of \cite{amsb_tata}) using superparticle
cascades of various kinds. For $m_{3/2} > 80~TeV$, such signals
have limited reach for $m_{0} > 1.2~TeV$. Our results, on the other hand,
essentially depend on direct chargino production and therefore are
independent of $m_{0}$. As can be seen from figures 2(a) and 2(b),
the overall reach of our suggested signal in $m_{3/2}$-space, even
at the $5\sigma$ level, is also higher.

We now come to the situation where the gaugino masses are large
compared to the $\mu$-parameter in a SUSY GUT. Treating $\mu$ as a
free parameter, one can have as small a separation as about 5 $GeV$
between $\chi^{\pm}_1$ and $\chi^{0}_1$ in such cases, while
$\chi^{0}_2$ can be within about 15 $GeV$ of $\chi^{\pm}_1$, so long
as one is within the region allowed by the LEP data. The branching
ratio for such a $\chi^{\pm}_1$ going to a pion is within about 10 per
cent. Invisibility of the chargino mostly hinges on the ensuing
leptons and quarks produced in three-body decays being sufficiently
soft. Very similar considerations apply to the decay of the
$\chi^{0}_2$ as well.

Since both of the two lightest neutralinos and the lighter chargino
are Higgsino-dominated in this case, the production rates for
$\chi^{\pm}_1 \chi^{0}_1$, $\chi^{\pm}_1 \chi^{0}_2$ and $\chi^{+}_1
\chi^{-}_1$ in gauge boson fusion are of comparable magnitudes. Thus
the computation of the $forward~jets + {E_T}\slash~$ has to take into
account all the above channels, with the three-body decay products
sufficiently degraded to escape detection.

We have used similar criteria for tagging the forward jets as in the
AMSB case, and the same ${E_T}\slash~$ and $\phi_{jj}$ cuts, with the
additional demand that the $E_T$ of jets, leptons or pions in the
central region be less than $10~GeV$. Events are seen to pass this
criteria only if the mass difference between $\chi^{\pm}_1$ and
$\chi^{0}_1$ is less than $10~GeV$. This region has been identified in
figures 3(a) and 3(b) with  $2~\sigma$ exclusion 
contours. The relative dilution of the results
compared to the AMSB scenario is due to the fact that only a finite
fraction of the three-body decay products are really soft enough to be
undetected, unlike the very soft pions that are inexorably produced in
the AMSB scenario. Also, since  $\chi^{\pm}_1$ and $\chi^{0}_1$
here are Higgsino-like, their gauge couplings are smaller than 
their gaugino-like counterparts which belong to the adjoint 
representation of $SU(2)$. This causes a further reduction compared
to AMSB.

Squark and slepton masses on the order of $400~GeV$ have been assumed
in the above analysis. This prevents two-body decays of $\chi^{\pm}_1$
and $\chi^{0}_2$ over most of the parameter space involved here. If
the sfermions are light enough to allow such decays, $\chi^{0}_2$, for
example, can decay into a sneutrino and a neutrino, both of which can
be invisible. The exclusion of such decays from our calculation makes
our estimates conservative.

In conclusion, invisibly decaying charginos can be used to our
advantage at the LHC if one concentrates on the chargino-pair and
chargino-neutralino productions via vector boson fusion. Tagging of
the two forward jets, with no other visible particle in the final
state, will allow us to constrain a large
part of the parameter space in AMSB  at the $5 \sigma$ level,
and a still larger region at the $2\sigma$ level. 
Also, a SUSY GUT scenario where
the lighter neutralinos and chargino are Higgsino-dominated can be
subjected to similar, though less stringent, constraints. Thus the
vector boson fusion channel can offer a major improvement in the
search strategies for the chargino-neutralino sector of SUSY theories
where such particles turn out difficult to detect otherwise.

{\bf Acknowledgements:} The work of BM was partially supported by the
Board of Research in Nuclear Sciences, Government of India. 
\newcommand{\plb}[3]{{Phys. Lett.} {\bf B#1} #2 (#3)}                  %
\newcommand{\prl}[3]{Phys. Rev. Lett. {\bf #1} #2 (#3) }        %
\newcommand{\rmp}[3]{Rev. Mod.  Phys. {\bf #1} #2 (#3)}             %
\newcommand{\prep}[3]{Phys. Rep. {\bf #1} #2 (#3)}                   %
\newcommand{\rpp}[3]{Rep. Prog. Phys. {\bf #1} #2 (#3)}             %
\newcommand{\prd}[3]{Phys. Rev. {\bf D#1} #2 (#3)}                    %
\newcommand{\np}[3]{Nucl. Phys. {\bf B#1} #2 (#3)}                     %
\newcommand{\npbps}[3]{Nucl. Phys. B (Proc. Suppl.)
           {\bf #1} #2 (#3)}                                           %
\newcommand{\sci}[3]{Science {\bf #1} #2 (#3)}                 %
\newcommand{\zp}[3]{Z.~Phys. C{\bf#1} #2 (#3)}                 %
\newcommand{\mpla}[3]{Mod. Phys. Lett. {\bf A#1} #2 (#3)}             %
 \newcommand{\apj}[3]{ Astrophys. J.\/ {\bf #1} #2 (#3)}       %
\newcommand{\jhep}[2]{{Jour. High Energy Phys.\/} {\bf #1} (#2) }%
\newcommand{\astropp}[3]{Astropart. Phys. {\bf #1} #2 (#3)}            %
\newcommand{\ib}[3]{{ ibid.\/} {\bf #1} #2 (#3)}                    %
\newcommand{\nat}[3]{Nature (London) {\bf #1} #2 (#3)}         %
 \newcommand{\app}[3]{{ Acta Phys. Polon.   B\/}{\bf #1} #2 (#3)}%
\newcommand{\nuovocim}[3]{Nuovo Cim. {\bf C#1} #2 (#3)}         %
\newcommand{\yadfiz}[4]{Yad. Fiz. {\bf #1} #2 (#3);             %
Sov. J. Nucl.  Phys. {\bf #1} #3 (#4)]}               %
\newcommand{\jetp}[6]{{Zh. Eksp. Teor. Fiz.\/} {\bf #1} (#3) #2;
           {JETP } {\bf #4} (#6) #5}%
\newcommand{\philt}[3]{Phil. Trans. Roy. Soc. London A {\bf #1} #2
        (#3)}                                                          %
\newcommand{\hepph}[1]{hep--ph/#1}           %
\newcommand{\hepex}[1]{hep--ex/#1}           %
\newcommand{\astro}[1]{(astro--ph/#1)}         %


\begin{thebibliography}{99}

\bibitem{bjorken} J. D. Bjorken, \prd{47}{101}{1993}.
\bibitem{dawson}
R.N. Cahn, S. Dawson, \plb{136}{196}{1984}.
\bibitem{zeppenfeld}
R. Godbole, S. Rindani, Z. Phys. {\bf C36} 1987) 395.
D. Rainwater, D. Zeppenfeld, \jhep{9712:005}{1997};
D. Rainwater, D. Zeppenfeld, K. Hagiwara, \prd{59}{010437}{1999};
D. Rainwater, D. Zeppenfeld,  \prd{60}{113004}{1999}; T. Plehn,
D. Rainwater, D. Zeppenfeld, \prd{61}{093005}{2000}.
\bibitem{susy_search} 
See for example, M. Dittmar, \hepex{990742};
H. Baer, talk given at SUSY-2001 at Dubna, Russia:
{\tt http://susy.dubna.ru/doc/baer.pdf}.
\bibitem{hadq} W. Beenakker {\it et al.}, \prl{83}{3780}{1999};
K. Matchev and D. Pierce, \prd{60}{075004}{1999};
H. Baer {\it et al}, \prd{61}{095007}{2000}. 
\bibitem{photon_tag}
C.-H. Chen, M. Drees, J. Gunion, \prl{76}{2002}{1996}; A. Datta, S. Maity,
\prd{59}{055019}{1999}; A. Datta, S. Maity, \plb{513}{130}{2001}.
\bibitem{eboli} O. Eboli and D. Zeppenfeld, \plb{495}{147}{2000}.
\bibitem{our_pap} A. Datta, P. Konar, B. Mukhopadhyaya, \hepph{0109071}.
\bibitem{amsb}L. Randall, R. Sundrum, \np{557}{79}{1999}; T. Gherghetta, 
G. Giudice, J. Wells, \np{559}{27}{1999}.  
\bibitem{sourov} D. K. Ghosh, P. Roy, S. Roy, JHEP 0008:031,2000; 
D. K. Ghosh, A. Kundu, P. Roy, S. Roy; \prd{64}{115001}{2001}.
\bibitem{amsb_tata}H. Baer, J.K. Mizukoshi, X. Tata, \plb{488}{367}{2000};
F. Paige, J. Wells, \hepph{0001249}.
\bibitem{rain} D. Rainwater, Ph.D. thesis, hep-ph/9908378.
\bibitem{helas} K. Hagiwara, H. Murayama, I. Watanabe; KEK-91-11, (1992).
\bibitem{cteq} H. Lai et al., \prd{55}{1280}{1997}.
\bibitem{feng} J. L. Feng {\it et al.}, \prl{83}{1731}{1999}. 
\bibitem{lep_data} M. Acciari et al., L3 Collab., \plb{482}{31}{2000}.
\end{thebibliography}
\end{document}